\documentclass[a4paper,11pt]{article}
\usepackage{pos}

\newcommand{\snn}{$\sqrt{s_{\rm{NN}}}$}
\newcommand{\s}{$\sqrt{s}$}
\newcommand{\ycms}{$y_{\rm{c.m.s.}}$}
\newcommand{\ylab}{$y_{\rm{lab}}$}

\title{Heavy-flavour studies with a high-luminosity fixed-target experiment at the LHC}
\ShortTitle{Heavy-flavour studies at AFTER@LHC}

\author*[1]{B.~Trzeciak}

\author[2]{S.J.~Brodsky} 
\author[3]{G.~Cavoto} 
\author[4]{C.~Da~Silva}
\author[5]{M.G.~Echevarria}
\author[6]{E.G.~Ferreiro}
\author[7]{C.~Hadjidakis}
\author[8]{R.~Haque}
\author[7]{I.~H\v{r}ivn\'{a}\v{c}ov\'{a}}
\author[8]{D.~Kiko\l a}
\author[4]{A.~Klein}
\author[9]{A.~Kurepin}
\author[10]{A.~Kusina}
\author[7]{J.P.~Lansberg}
\author[11]{C.~Lorc\'e}
\author[12]{F.~Lyonnet}
\author[13]{Y.~Makdisi}
\author[7]{L.~Massacrier} 
\author[14]{S.~Porteboeuf}
\author[15]{C.~Quintans}
\author[16]{A.~Rakotozafindrabe} 
\author[7]{P.~Robbe}
\author[17]{W.~Scandale} 
\author[18]{I.~Schienbein}
\author[15,19]{J.~Seixas}
\author[20]{H.S.~Shao}
\author[21,22]{A.~Signori}
\author[9]{N.~Topilskaya}
\author[23]{A.~Uras} 
\author[7]{C.~Van~Hulse}
\author[24]{J.~Wagner}
\author[7]{N.~Yamanaka}
\author[25]{Z.~Yang}
\author[13]{A.~Zelenski}
\small
\affiliation[1]{Faculty of Nuclear Sciences and Physics Engineering, Czech Technical University in Prague, Brehova 7, 115 19 Prague, Czech Republic}
\affiliation[2]{SLAC National Accelerator Laboratory, Stanford University, Menlo Park, CA 94025, USA}
\affiliation[3]{``Sapienza" Universit\`a di Roma, Dipartimento di Fisica \& INFN, Sez. di Roma, P.le A. Moro 2, 00185 Roma, Italy}
\affiliation[4]{P-25, Los Alamos National Laboratory, Los Alamos, NM 87545, USA}
\affiliation[5]{Department of Physics and Mathematics, University of Alcal\'a, 28805 Alcal\'a de Henares (Madrid), Spain}
\affiliation[6]{Dept. de F{\'\i}sica de Part{\'\i}culas \& IGFAE, Universidade de Santiago de Compostela, 15782 Santiago de Compostela, Spain}
\affiliation[7]{Universit\'e Paris-Saclay, CNRS, IJCLab, 91405 Orsay, France}
\affiliation[8]{Faculty of Physics, Warsaw University of Technology, ul. Koszykowa 75, 00-662 Warsaw, Poland}
\affiliation[9]{Institute for Nuclear Research, Moscow, Russia}                    
\affiliation[10]{Institute of Nuclear Physics Polish Academy of Sciences, PL-31342 Krakow, Poland}
\affiliation[11]{CPHT, CNRS, Ecole Polytechnique, Institut Polytechnique de Paris, 91128 Palaiseau, France}
\affiliation[12]{Southern Methodist University, Dallas, TX 75275, USA}
\affiliation[13]{Brookhaven National Laboratory, Collider Accelerator Department}
\affiliation[14]{Universit\'e Clermont Auvergne, CNRS/IN2P3, LPC, F-63000 Clermont-Ferrand, France.}
\affiliation[15]{LIP, Av. Prof. Gama Pinto, 2, 1649-003 Lisboa,Portugal}
\affiliation[16]{IRFU/DPhN, CEA Saclay, 91191 Gif-sur-Yvette Cedex, France}
\affiliation[17]{CERN, European Organization for Nuclear Research, 1211 Geneva 23, Switzerland}
\affiliation[18]{Laboratoire de Physique Subatomique et de Cosmologie, Universit\'e Grenoble Alpes, CNRS/IN2P3, 
53 Avenue des Martyrs, F-38026 Grenoble, France}
\affiliation[19]{Dep. Fisica, Instituto Superior Tecnico, Av. Rovisco Pais 1, 1049-001 Lisboa, Portugal}
\affiliation[20]{LPTHE, UMR 7589, Sorbonne University\'e et CNRS, 4 place Jussieu, 75252 Paris Cedex 05, France}
\affiliation[21]{Theory Center, Thomas Jefferson National Accelerator Facility, 12000 Jefferson Avenue, Newport News, VA 23606, USA}
\affiliation[22]{Dipartimento di Fisica, Universit\`a di Pavia, via Bassi 6, I-27100 Pavia, Italy}
\affiliation[23]{IPNL, Universit\'e Claude Bernard Lyon-I and CNRS-IN2P3, Villeurbanne, France}
\affiliation[24]{National Centre for Nuclear Research (NCBJ), Ho\.{z}a 69, 00-681, Warsaw, Poland}
\affiliation[25]{Center for High Energy Physics, Department of Engineering Physics, Tsinghua University, Beijing, China}

\emailAdd{barbaraAntonina.Trzeciak@fjfi.cvut.cz}

\abstract{
Extraction of the multi-TeV proton and lead LHC beams with a bent crystal or by using an internal gas target allows one to perform the most energetic fixed-target experiment ever. $pp$, pd and $p$A collisions at \snn~=~115~GeV and Pb$p$ and PbA collisions at \snn~=~72~GeV can be studied with high precision and modern detection techniques over a broad rapidity range.
Using the LHCb or the ALICE detector in a fixed-target mode offers unprecedented possibilities to access heavy-flavour production in a new energy domain, half way between the SPS and the nominal RHIC energy.
In this contribution, a review of projection studies for quarkonium and open charm and beauty production with both detector set-ups used with various nuclear targets and the LHC lead beams is presented.
}

\FullConference{%
  HardProbes2020\\
  1-5 June 2020\\
  Austin, Texas}

\begin{document}
\maketitle

\vspace{-25pt}
\section{Introduction and Physics Motivations}
\vspace{-10pt}

A fixed-target mode with the 7~TeV LHC proton and 2.76~A~TeV lead beam provides the center-of-mass (c.m.s.) energy per nucleon pair of \snn~=~115~GeV for $pp$, pd and $p$A and \snn~=~72~GeV for Pb$p$ and PbA collisions, and the c.m.s. rapidity boost of 4.8 and 4.2 units, respectively. It offers variety of polarised and nuclear targets, and high luminosities thanks to dense and long targets. Due to the large rapidity boost, the backward rapidity region (\ycms~$<$~0) and so the high-$x$ domain ($x$ is the momentum fraction of the parton in the target nucleon) are accessible with the standard experimental techniques.

A fixed-target experiment utilizing the multi-TeV LHC proton and lead beams would greatly complement the current LHC collider programs, in the new energy domain between the SPS and the nominal RHIC collider energy.
The main physics motivations for a fixed-target experiment at LHC are threefold: (i) the high momentum fraction ($x$) frontier in nucleons and nuclei (ii) the spin content of the nucleons (iii) studies of the hot and dense medium created in ultra-relativistic heavy-ion collisions down to the target rapidity region ~\cite{Brodsky:2012vg,Hadjidakis:2018ifr}.
In this contribution we will focus on the physics cases accessible with heavy-flavour probes. More details on the physics program, as well as a technical review of the project can be found in~\cite{Hadjidakis:2018ifr,Barschel:2020drr}.

Production of the open heavy-flavour hadrons and quarkonia is sensitive to the gluon content of the proton. A fixed-target experiment at LHC would allow one to constrain the gluon parton distribution function (PDF) in the region of the high-momentum fractions (high-$x$) where the uncertainties are large~\cite{Kusina:2019grp}. 
Moreover, at the high-$x$ domain the predicted non-perturbative intrinsic source of the heavy-quark contribution of the nucleon~\cite{Sufian:2020coz} can influence the open heavy-flavour production~\cite{Brodsky:2015fna}. This contribution can be verified with the proposed fixed-target experiment. These studies are also of particular importance for the high-energy neutrino and cosmic ray physics. 
Transversely polarised targets allow one to measure the Single Transverse Spin Asymmetries and in this way to access the gluon Sivers function in order to study the gluon orbital angular momentum in the proton~\cite{Kikola:2017hnp}.
A fixed-target program with a proton beam on different nuclear targets enables extensive studies of the Cold Nuclear Matter (CNM) effects, such as modifications of the parton distributions in the nuclear medium, energy loss or nuclear absorption. 
In particular, heavy-flavour probes give an access to the gluon nuclear PDF which at high-x would allow one to study the still very poorly known gluon EMC effect.
Good understanding of the CNM effects from $p$A collisions would also help in more precise studies of the properties of the deconfined hot and dense medium~- Quark Gluon Plasma (QGP)~- expected to be created in PbA collisions at \snn~=~72~GeV. The wide rapidity coverage of such an experiment would allow one to access the longitudinal expansion of the medium and the phase diagram of the nuclear matter at different values of $\mu_{B}$~\cite{Karpenko:2018xam,Begun:2018efg} which would be a complementary approach to the RHIC Beam Energy Scan programs. 
Hard probes are of particular interest here as they can provide a good handle on the dynamic and thermodynamic properties of the medium, especially at mid-rapidity (\ycms~$\approx$~0) where the yields are the highest.
Heavy quarks are created at the very early stage of the heavy-ion collisions in the initial hard scatterings and participate in the whole medium evolution loosing their energy via radiative and collisional interactions with the medium. Studies of the sequential suppression of different bottomonium states, $\Upsilon(1S,2S,3S)$, have a potential to serve as the {\it medium thermometer} -- with all the caveats related to the dynamical effects beyond the static Debye-like screening of the $Q\bar{Q}$ potential in the coloured medium, which prevent the usage of the charmonium states as the thermometer. 

\vspace{-15pt}
\section{Possible implementations}
\vspace{-10pt}

There are several technical implementations that can be considered in order to realise a fixed-target experiment with the LHC proton and lead beams~\cite{Hadjidakis:2018ifr}. 
The most promising and feasible solutions are: (i) an internal gas target (ii) a bent crystal splitting the beam halo onto an internal gas or solid target -- both options can be installed in the existing LHCb or ALICE experiments. Feasibility of the possibility (i), however with low gas densities and without the target polarisation, was already demonstrated by the SMOG system at LHCb~\cite{FerroLuzzi:2005em} and the recently installed SMOG2 system~\cite{LHCbCollaboration:2673690}.

\vspace{-15pt}
\section{Projection studies}
\vspace{-10pt}

This contribution focuses on heavy-flavour projection studies with a fixed-target experiment at the LHC using unpolarised targets. 
In particular, we present new studies of $D^{0}$ (and $\bar{D^{0}}$) meson production with an ALICE-like detector setup. This setup assumes a 1-cm long solid target located -4.7m away from the ALICE nominal interaction point (ALICE A-side). Hadronic $D^{0} \rightarrow K^{-}\pi^{+}$ decays are considered and $D^{0}$ mesons are reconstructed from the decay products identified in the ALICE central barrel which provides a backward rapidity coverage for $D^{0}$, -3.5~$<$~\ycms~$<$~-2.3.
The simulations were performed using the PYTHIA8 Monash2013 tune with $hardQCD$ processes, and the total $c\bar{c}$  production cross section was scaled to 2.29$\times 10^{-1}$~$mb$ based on the HELAC-ONIA~\cite{Shao:2012iz,Shao:2015vga} studies in $pp$ collisions at \s~=~115~GeV~\cite{Massacrier:2015qba}.
Precise $D^{0}$ measurements would also require an additional vertex detector placed close to the target position in order to reconstruct the secondary vertex of the $D^{0}$ meson decays.
The detector response was parametrised -- the assumed $D^{0}$ efficiency $\times$ acceptance is 2\% (based on the LHCb SMOG performance~\cite{LHCb-CONF-2017-001}) and the second order event plane ($\Psi_2$) resolution is taken to be 20\%. Three nuclear target cases were studied, they would provide the yearly integrated luminosities of 1.12~$pb^{-1}$ for $p$C, 0.56~$pb^{-1}$ for $p$Ti and 0.64~$pb^{-1}$ for $p$W systems.
	
	\vspace{-10pt}
\begin{figure}[!htb]
 {\includegraphics[width=0.44\textwidth]{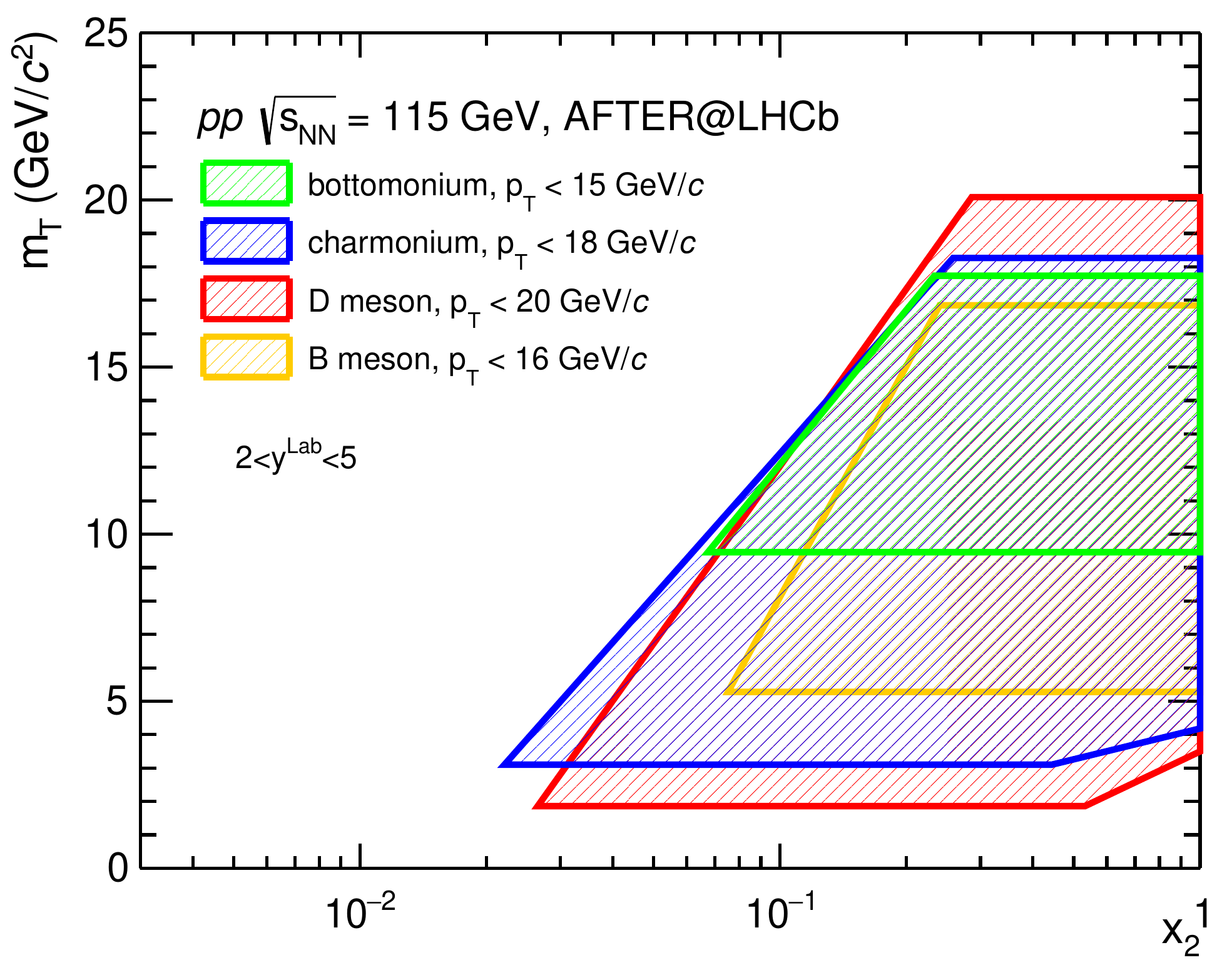}} \hspace{20pt}
 {\includegraphics[width=0.5\textwidth]{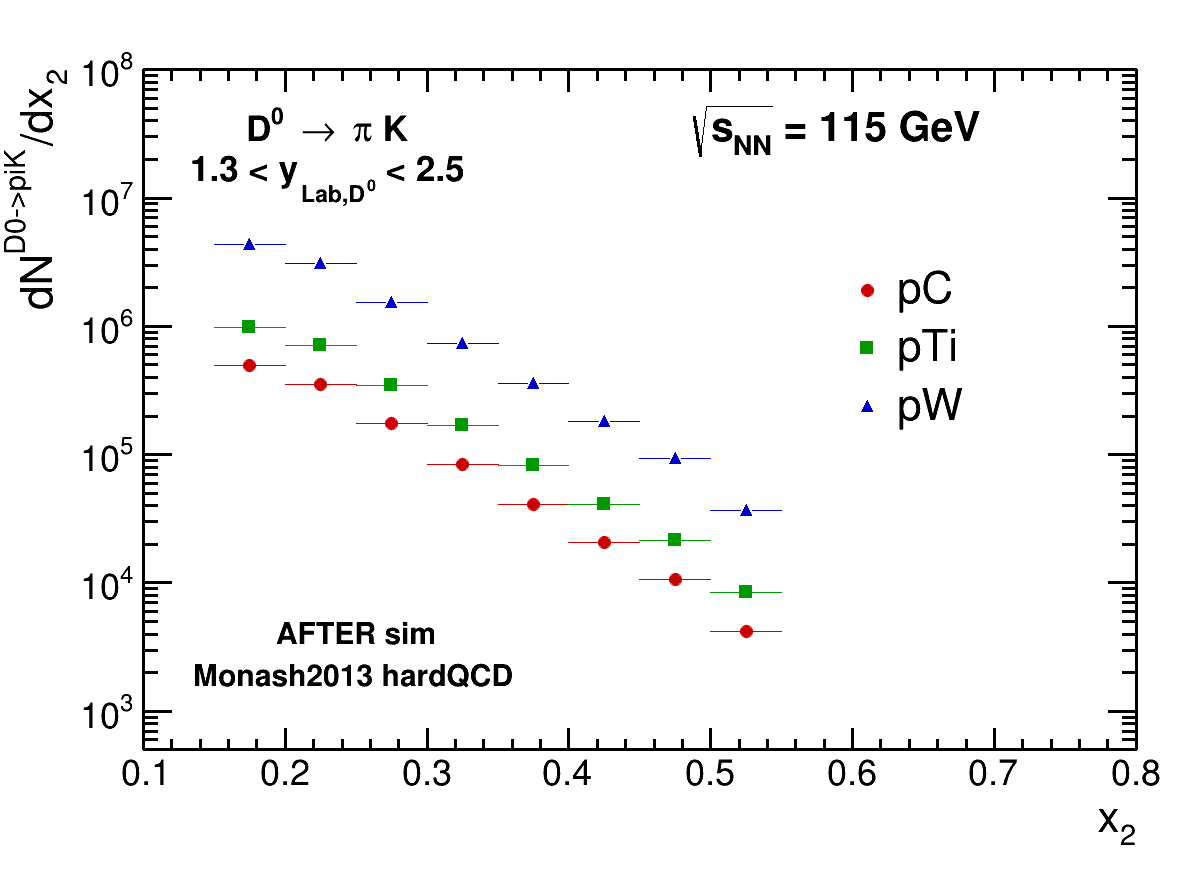}}
\caption{Left: Typical kinematical reach in $x_{2}$ with bottomonium, charmonium and D, B meson probes and at $m_{T}$ scale of the fixed-target mode with the LHCb-like detector acceptance in $pp$ collisions at \s~=~115~GeV. Right: Expected $D^{0}$ meson yields with the ALICE-like setup as a function of $x_{2}$, with the proton beam on different nuclear targets at \snn~=~115~GeV, for one LHC running year.}
\label{fig:DmesonYield}
\end{figure}
\vspace{-12pt}

The presented projections based on a LHCb-like detector consider gaseous H and Xe targets placed in the LHCb nominal interaction point. The yearly integrated luminosities of 10~$fb^{-1}$, 100~$pb^{-1}$ and 30~$nb^{-1}$ for the $pp$, $p$Xe and PbXe collisions, respectively, are considered. Details of these simulations are reported in~\cite{Massacrier:2015qba,Trzeciak:2017csa,Hadjidakis:2018ifr}.

\begin{figure}[!htb]
\vspace{-10pt}
 {\includegraphics[width=0.4\textwidth]{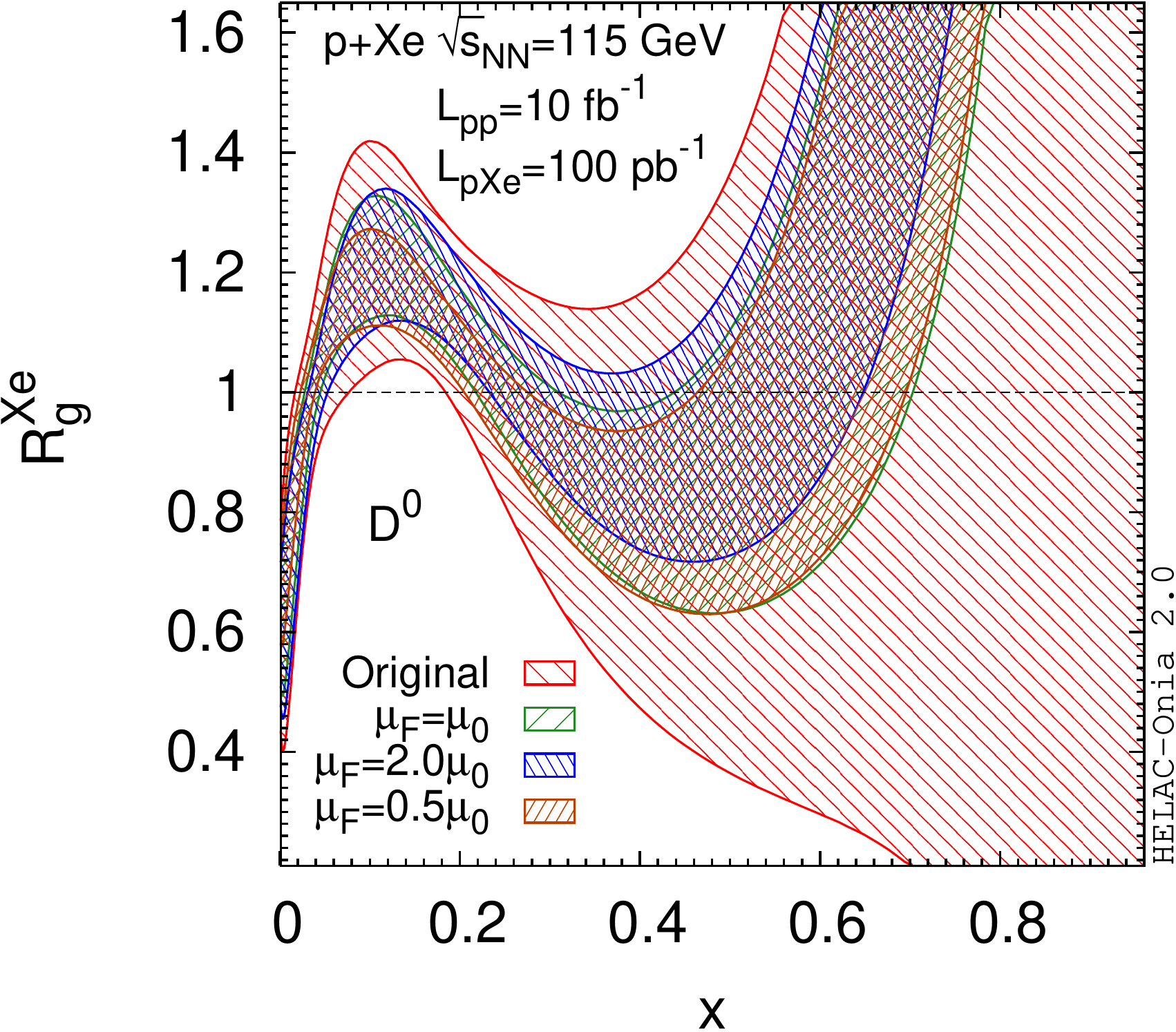}} \hspace{30pt}
  {\includegraphics[width=0.45\textwidth]{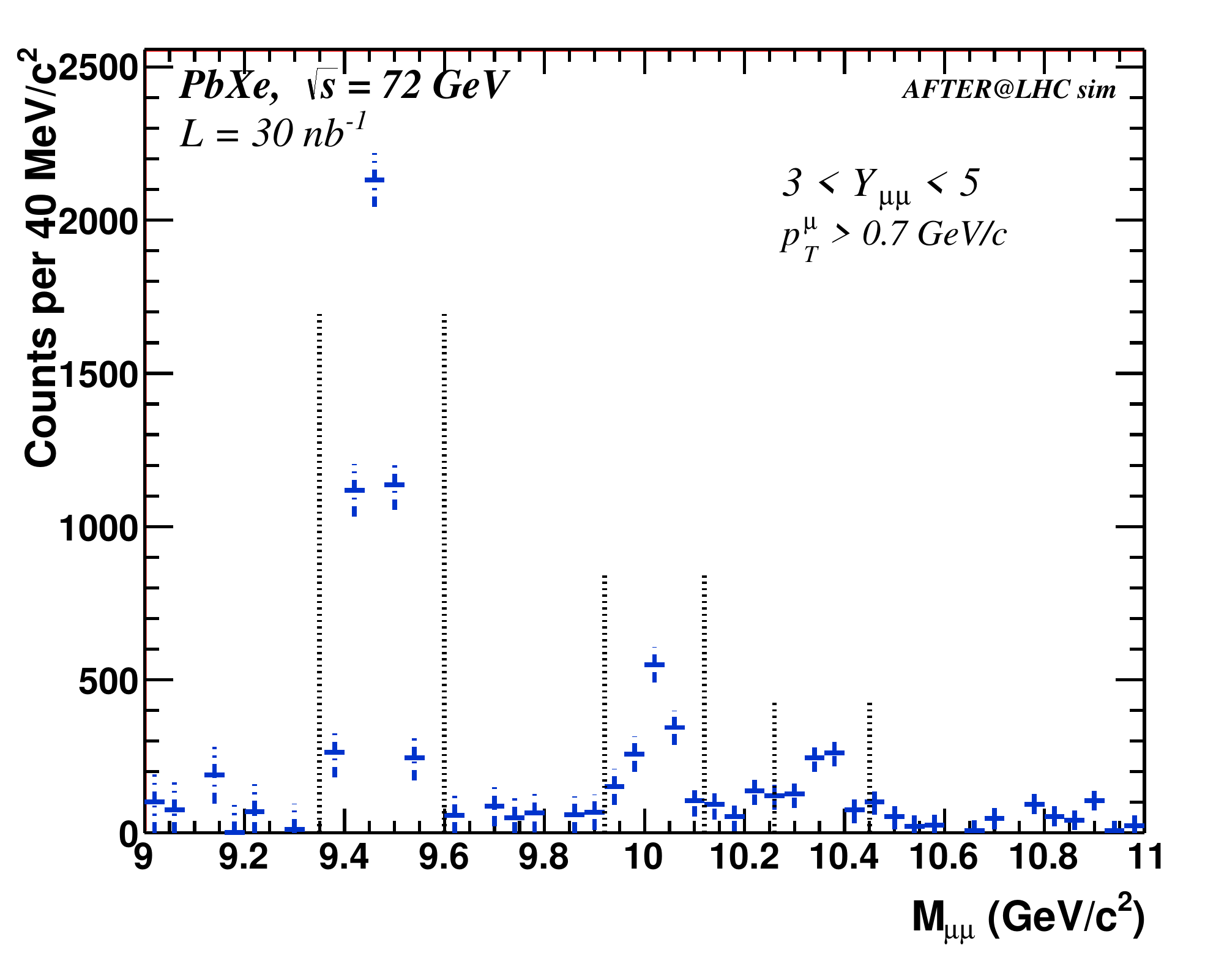}}
\caption{Left: nCTEQ15 gluon nPDFs (ratio of gluon densities in nCTEQ15/CT14 PDFs) before and after the reweighting using fixed-target $D^{0}$ $R_{p\rm{Xe}}$ pseudo-data from LHCb-like simulations at a scale $Q=$~2~GeV. 
Right: The di-muon invariant mass distribution (the uncorrelated background subtraction) in the $\Upsilon(nS)$ mass range, PbXe collisions at \snn~=~72~GeV with the LHCb-like setup. No nuclear modifications are assumed.}
\label{fig:LHCb}
\end{figure}

Figure~\ref{fig:DmesonYield} left presents a kinematical reach in $x_{2}$ for heavy-flavour hadron production with the scale chosen to be $m_{T}$ ($m_{T} = \sqrt{M^{2}_{\rm{hadron}} + p^{2}_{\rm{T, hadron}}}$), in $pp$ (or $p$A) collisions at \s~=~115~GeV with the LHCb-like acceptance in the laboratory frame of 2~$<$~\ylab~$<$~5.
The expected yearly $D^{0}$ yields (the same yields are expected for $\bar{D^{0}}$) in $p$A collisions at \snn~=~115~GeV with the ALICE-like detector configuration as a function of $D^{0}$ $x_{2}$ are shown in Fig.~\ref{fig:DmesonYield} right. The $D^{0}$ rapidity acceptance is 1.3~$<$~\ylab~$<$~2.5 and the obtained $p_{T}$ interval is 0--5~GeV/$c$.
An example of the impact of heavy-flavour studies on the gluon nPDF can be seen in Fig.~\ref{fig:LHCb} left, where pseudo-data from $pp$ and $p$Xe collisions with LHCb-like detector setup were used. Only with $D^{0}$ $R_{p\rm{Xe}}$ pseudo-data a large improvement on the gluon nPDF uncertainties, especially at high $x$, is clearly visible.

\vspace{-12pt}

\begin{figure}[!htb]
\vspace{-10pt}
 {\includegraphics[width=0.5\textwidth]{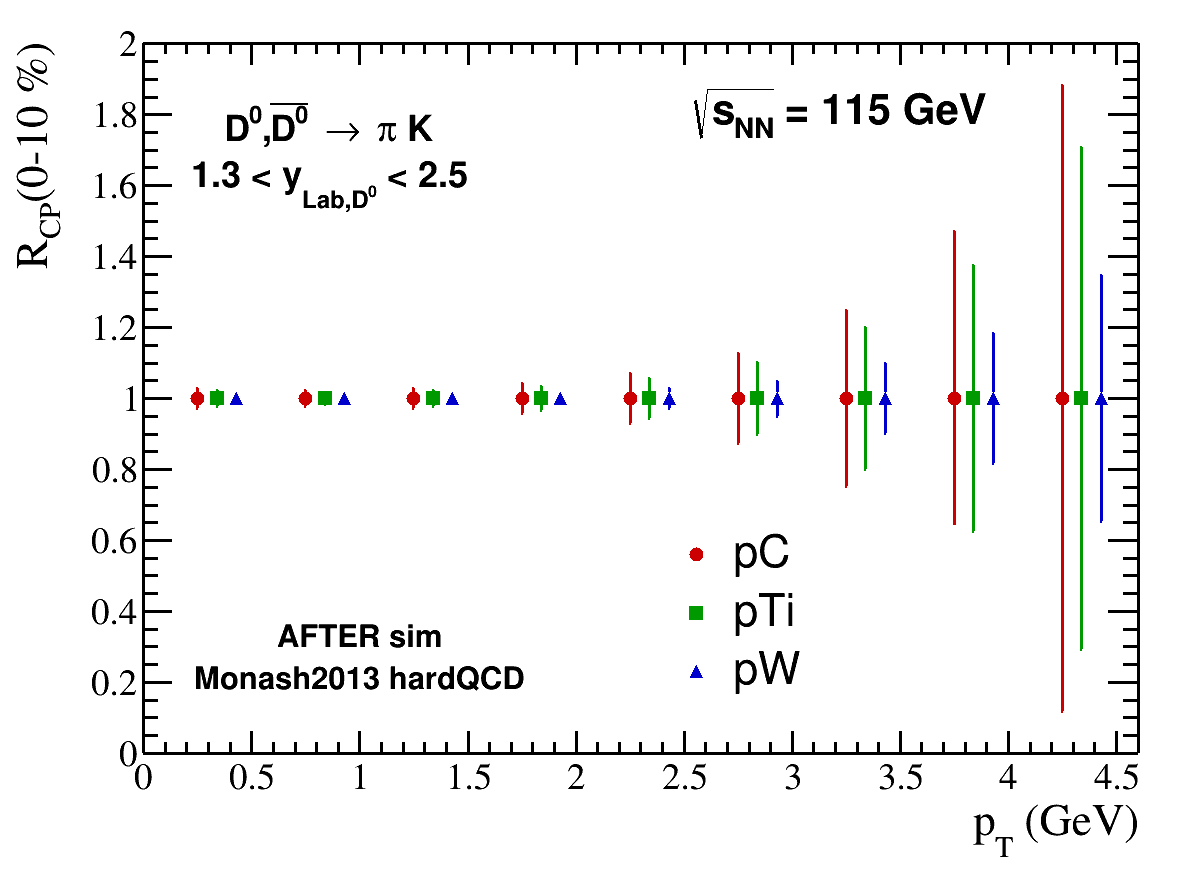}}
  {\includegraphics[width=0.5\textwidth]{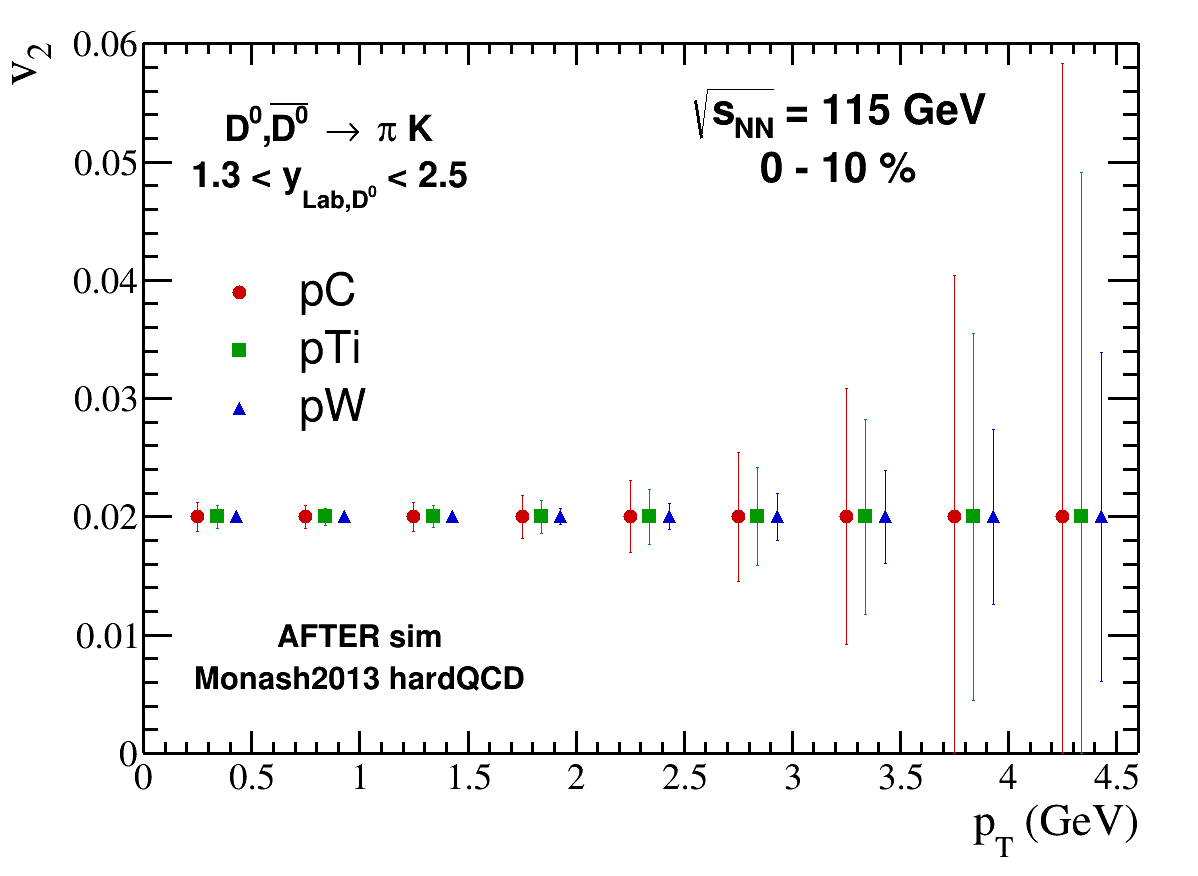}}
\caption{Statistical projections on $R_{CP}$ (left) and $v_2$ (right) vs $D^{0}$ $p_{T}$ in 0--10\% central $p$A collisions at \snn~=~115~GeV for the ALICE-like detector setup.}
\label{fig:DmesonRcp}
\vspace{-15pt}
\end{figure}

Figure~\ref{fig:DmesonRcp} shows statistical projections on $R_{CP}$ (nuclear modification factor in central collisions w.r.t. 60--100\% peripheral collisions) and the elliptic flow $v_2$ coefficient (the second order flow harmonic of the Fourier expansion of the azimuthal distribution of the produced hadrons) as a function of $D^{0}$ $p_{T}$ in 0--10\% central $p$A collisions at \snn~=~115~GeV. The projections are based on the expected $D^{0}$ and $\bar{D^{0}}$ yearly yields with the ALICE-like detector setup and number of events per 10\%-wide centrality class: 15$\times 10^{9}$, 20$\times 10^{9}$ and 80$\times 10^{9}$ for $p$C, $p$Ti and $p$W systems, respectively.
These measurements would not only serve a CNM baseline for heavy-ion studies, but would also provide an important input for further investigation of the collective-like effects in small systems that have been observed by the LHC and RHIC experiments, their origin is however still not understood. Extensive heavy-flavour studies can also be performed in heavy-ion collisions with the LHC lead beam on different nuclear targets. An example of $\Upsilon(nS)$ signal reconstruction using the LHCb-like detector setup is shown in Fig.~\ref{fig:LHCb} right. With the excellent LHCb momentum resolution it would be possible to clearly separate the different $\Upsilon(nS)$ states and thanks to the high enough yields to perform measurements of the $\Upsilon(nS)$ nuclear modification factors.

Moreover, heavy-flavour studies in ALICE can be extended with analyses of quarkonia via the di-muon decay channel or using muons from the open heavy-flavour hadron decays thanks to the ALICE muon detectors.
These measurements would give an access to the heavy-flavour production in the c.m.s. rapidity region of -1.6~$<$~\ycms~$<$~-0.5. In addition, with the high expected yields one could also perform, e.g., high precision heavy-flavour $v_{1}$ or correlation studies. 
More detailed projection studies, with a full ALICE detector simulation and the LHC beam halo deflected by a bent crystal on different solid targets, are currently ongoing. 

\vspace{-12pt}
\section{Summary} 
\vspace{-10pt}
In these proceedings, we have briefly discussed the main physics cases and presented projection studies for a fixed-target experiment with the multi-TeV LHC proton and lead beams, focusing on heavy-flavour probes. 
In particular, new $D^{0}$ meson projections with an ALICE-like detector setup were presented. 

\vspace{-12pt}
\acknowledgments 
\vspace{-10pt}
{\small
BT acknowledges support from The Czech Science Foundation, grant number: GJ20-16256Y. This project has received funding from the European Union’s Horizon 2020 research and innovation programme under grant agreement No 824093. This research was also supported by the RFBR/CNRS grant 18-52-15007.
}

\small
\vspace{-8pt}
\bibliographystyle{h-elsevier}
{\linespread{0.1}\selectfont\bibliography{bib}}

\end{document}